\documentclass[10pt,twocolumn,twoside]{article}

\usepackage{mathtools, cuted}
\usepackage{lipsum, color}
\usepackage[export]{adjustbox}
\usepackage{graphicx}
\usepackage{float}
\usepackage{caption}
\usepackage{subcaption}
\usepackage{empheq}
\usepackage{setspace}
\usepackage[T1]{fontenc}
\usepackage[utf8]{inputenc}
\usepackage{authblk}
\usepackage{widetext}

\usepackage[utf8]{inputenc}
\usepackage[english]{babel}
\usepackage{amsmath,amsfonts,amssymb}
\usepackage{mathpazo} % Widely available alternative to Garamond
\usepackage[scaled]{helvet}
\usepackage[T1]{fontenc}
\usepackage{url}

\title{Mode analysis of higher-order transverse-mode correlation beams in turbulent atmosphere}
\author[1,2]{H. Avetisyan}
\author[1,*]{C.H. Monken}
\affil[1]{Departamento de F\'isica, Universidade Federal de Minas Gerais, Caixa Postal 702, \\ Belo Horizonte, MG 30161-970, Brazil}
\affil[2]{Institute for Physical Research, Armenian National Academy of Sciences, Ashtarak-2, 0203, Armenia}

\affil[*]{Corresponding author: monken@fisica.ufmg.br}

\begin{document}
\maketitle
\begin{abstract}\textbf{
Due to the transfer of the angular spectrum of the pump beam to the two-photon state in spontaneous parametric down-conversion the generated twin photons are entangled in Hermite-Gaussian (HG) modes. This enables one using HG modes as an alphabet for quantum communication. For global quantum communication purposes, we derive an analytic expression for two-photon detection probability in terms of HG modes taking into account the effects of the turbulent atmosphere. Our result is more general as it accounts for the propagation of both, signal and idler photons through the atmosphere, as opposed to other works considering one of the photons' propagation in vacuum. 
We show that while the restrictions on both the parity and order of the down-converted HG fields no longer hold due to the crosstalk between modes when propagating in the atmosphere, the crosstalk is not uniform: there are more robust modes that tend to keep the photons in them. These modes can be employed in order to increase the fidelity of quantum communication.}
\end{abstract}

%\setboolean{displaycopyright}{true}

%\thispagestyle{fancy}

%\ifthenelse{\boolean{shortarticle}}{\ifthenelse{\boolean{singlecolumn}}{\abscontentformatted}{\abscontent}}{}

\section{Introduction}

The fact that two-photon states generated by spontaneous parametric down-conversion (SPDC) are entangled in transverse spatial modes with high Schmidt numbers  opens a possibility of encoding information in two-photon states using larger alphabets \cite{Walborn2010,DiLorenzo2009}. To this end, it is of importance to understand whether the transverse mode correlation is still present in any extent after the down-converted two-photon states have propagated through a turbulent medium.
\newline
Among the many possible sets of orthogonal higher-order optical beam modes, the best known are the Laguerre-Gaussian (LG) and HG modes. LG modes are of particular interest as their helical phase front structures carry orbital angular momentum (OAM)\cite{Allen1992}. Due to this complex spatial distribution, LG modes cannot be efficiently coupled to single-mode fibers. In order to detect higher-order modes, computer generated holograms shall be used to transform the higher-order modes to the zero-order ones which are further coupled to single-mode fibers \cite{Arlt1998}. Computer generated holograms can also be used to project superposition states of LG modes to a particular state defined by the hologram necessary for verification of entanglement \cite{Mair, Vaziri2002,Vaziri2003}. As the hologram and the single mode fiber configuration is sensitive to the radial distribution of the source (characterized by the mode number $p$), measurements of only the spiral spectrum of entangled two-photon states have been carried out \cite{DiLorenzo2010}.
%This is the main reason why the previous works could not be directly compared with predictions of the well-known SPDC wave function.
\newline
In the presence of turbulence, the evolution of entanglement in the three dimensional LG mode basis has been observed using a single phase screen model for one of the photons, either signal or idler \cite{arXiv:1604.06237}. Recently, an experiment on transmission of OAM modes of light over a distance of 143 kilometers has been performed \cite{Malik}.
\newline
The atmospheric fourth-order correlation function for the SPDC process, when the field of pump represents any of HG or
LG modes as well as a partially coherent field has been calculated analytically in Ref. \cite{Avetisyan2016},
showing that the joint probability for detection of the signal and idler photons in different positions
is considerable after propagation for more than five kilometers.
In this Letter, we go a step forward and calculate the joint probability to detect the photons in different HG modes (see Eq. (\ref{final_expression})) which is a product of functions, which mixes the indices of signal and idler implying that the entanglement is preserved. Our results also show that some pairs of modes are more robust to the crosstalk due to atmospheric propagation than others which can be beneficial for free space quantum communication purposes.
\smallskip
\section{Probability of Detection of Transverse-Mode Correlation Beams}
Full spatial entanglement has been accessed experimentally with feasible radial detection modes with negligible cross correlations
in vacuum \cite{Salakhutdinov,Krenn2013}.
The expected perfectly (anti)correlated pure state from SPDC has the form
\begingroup\makeatletter\def\f@size{8}\check@mathfonts
\def\maketag@@@#1{\hbox{\m@th\large\normalfont#1}}
\begin{equation}
\vert \psi \rangle=  \sum_{p=0}^{\infty}\sum_{l=-\infty}^{\infty}a_{p,l}\vert LG_{p,l}\rangle_s\vert LG_{p,-l}\rangle_i,
\end{equation}
where the coefficients $a_{p,l}$ are the probability amplitudes to detect a signal photon in the $(p,l)$ mode and an idler photon in the $(p,-l)$ mode.
In contrast to the azimuthal modes, their radial counterparts do not necessarily represent Schmidt modes \cite{Law&Eberly2004}, however, there has been found non-zero quantum correlations of detected modes with different $p$ \cite{Salakhutdinov}.
Instead of the basis of the LG modes, mathematically, it is less costly to make use of the HG modes.
As the HG modes form a complete set, one can expand the two-photon state as
\begin{equation}
\vert \psi_{mn}\rangle=\sum\limits_{j,k,u,t=0}^\infty C_{jkut}^{mn}\vert HG_{jk}\rangle_s\vert HG_{ut}\rangle_i,
\end{equation}
where $\vert \psi_{mn}\rangle$ is the state prepared in SPDC process and
$%\begin{equation}
C_{jkut}^{mn}= \null_s\langle HG_{jk}\vert \null_i\langle HG_{ut}\vert\psi_{mn}\rangle$
%\end{equation}
are the coefficients of expansion representing the probability amplitudes of detecting signal and idler photons
in HG modes with mode indexes $jk$ and $ut$, respectively.
The joint detection probability, $|C_{jkut}^{mn}|^2$,
for signal and idler photons each in some transverse mode propagating in vacuum has
been calculated before \cite{Walborn2004, Walborn2005}.
Following Ref.\cite{Franke-Arnold}, we write the joint probability for two photons in modes $\mathrm{M}_1$ and $\mathrm{M}_2$,
in the representation of the configuration space variables which facilitates accounting for the atmospheric effects on the state
\begin{equation}\label{prob_transverse_modes}
\begin{split}
P(M_1&,M_2)=\left |\langle\psi_1,\psi_2\vert\psi\rangle\right |^2 \\& \propto
\left |\int d\textbf{x}_1\int d\textbf{x}_2 M^{\ast}_1(\textbf{x}_1)M^{\ast}_2(\textbf{x}_2)
E_p \left( \dfrac{\textbf{x}_1+\textbf{x}_2}{2}\right)V(\textbf{x}_1-\textbf{x}_2) \right |^2,
\end{split}
\end{equation}
where
\begin{equation}
\vert\psi\rangle\propto \iint d\textbf{x}_1 d\textbf{x}_2
M\left( \frac{\textbf{x}_1+\textbf{x}_2}{2}\right)V(\textbf{x}_1-\textbf{x}_2)
\hat{a}^{\dagger}_1(\textbf{x}_1)\hat{a}^{\dagger}_2(\textbf{x}_2)\vert 0\rangle
\end{equation}
is the two-photon state generated by SPDC \cite{Monken98}, while
$E_p \left( \dfrac{\textbf{x}_1+\textbf{x}_2}{2}\right)V(\textbf{x}_1-\textbf{x}_2)$ represents the two photon wavefunction.
Experimentally, the modes $M_1(\textbf{x}_1)$ and $M_2(\textbf{x}_2)$ represent phase holograms, say,
to be coupled with a detection system. The frequency degenerate SPDC state is of our specific interest, that is, $\omega_s=\omega_i=\omega_p/2.$
\newline
Expression (\ref{prob_transverse_modes})
%(\ref{prob_transverse_modes1})
is independent of positions of detectors due to the fact that the fields, whose modal expansion is
made up of arbitrary weighted HG modes, are shape-invariant \cite{Gori}. The shape-invariant
property will no longer hold due to distortions caused by turbulence.
The two photon wavefunction $E_p \left( (\textbf{x}_1+\textbf{x}_2)/2)\right)V(\textbf{x}_1-\textbf{x}_2)$
thus can be taken in the far field approximation important for calculating the probability (\ref{prob_transverse_modes}) by
taking the effects of turbulence into account.

\section{Detection probability in the presence of turbulence.}

In this section we calculate the two-mode joint detection probability  (\ref{prob_transverse_modes}) taking into account
the effects of turbulence. To do so, we write the two-photon
wavefunction in the form \cite{Pereira}
\begin{equation}\label{2-phot-wf}
\begin{split}
E_p \left( (\textbf{x}_1+\textbf{x}_2)/2)\right)&V(\textbf{x}_1-\textbf{x}_2)=
\frac{1}{\lambda^2 z^2}\int d\textbf{r}'\int d\textbf{r}'' E_p \left( \dfrac{\textbf{r}'+\textbf{r}''}{2}\right)\\&\times
\delta (\textbf{r}'-\textbf{r}'')
\exp\left[\frac{ik}{2z}\left[ \vert\textbf{x}_1-\textbf{r}'\vert^2 +\vert\textbf{x}_2-\textbf{r}''\vert^2 \right]\right]\\&\times
\exp[\psi (\textbf{x}_1,\textbf{r}')+\psi (\textbf{x}_2,\textbf{r}'')],
\end{split}
\end{equation}
where $\psi (\textbf{x},\textbf{r})$ is a random function representing phase and amplitude
distortions of signal and idler fields, $\lambda$ is the wavelength and $z$ is the propagation distance.
Utilizing (\ref{2-phot-wf}), the probability (\ref{prob_transverse_modes})
takes the form
\begin{equation}\label{prob_trans_modes1}
\begin{split}
P(M_1&,M_2)=\mathcal{C}_0\int d\textbf{x}_1\int d\textbf{x}'_1 \int d\textbf{x}_2 \int d\textbf{x}'_2
\int d\textbf{r}' \int d\textbf{r}'' \\&\times
M^{\ast}_1(\textbf{x}_1)M_1(\textbf{x}'_1)
M^{\ast}_2(\textbf{x}_2)M_2(\textbf{x}'_2)
E_p (\textbf{r}')E^{\ast}_p (\textbf{r}'')\\&\times
\exp\left[\frac{ik}{2z}\left(\vert\textbf{x}_1-\textbf{r}'\vert^2 -
\vert\textbf{x}'_1-\textbf{r}''\vert^2+\vert\textbf{x}_2-\textbf{r}'\vert^2-
\vert\textbf{x}'_2-\textbf{r}''\vert^2\right) \right] \\&\times
\left\langle \exp\left[\psi(\textbf{x}_1, \textbf{r}')+\psi^{\ast}(\textbf{x}'_1, \textbf{r}'')+
\psi(\textbf{x}_2, \textbf{r}')+\psi^{\ast}(\textbf{x}'_2, \textbf{r}'') \right]\right\rangle,
\end{split}
\end{equation}
where $\mathcal{C}_0=1/(\lambda^4 z^4).$
%The integration for a Gaussian pump and HG mode functions is provided in the Supplementary Material, where the geometrical optics approximation along with the stochastic function $\psi(\textbf{r}_1,\textbf{r}_2)$ as a Gaussian random field were used. The result is
The integration for a Gaussian pump and HG mode functions, where the geometrical optics approximation along with the stochastic function $\psi(\textbf{r}_1,\textbf{r}_2)$ as a Gaussian random field were used, yields
\begin{equation}\label{final_expression}
P(HG_{m_sn_s}, HG_{m_in_i}) = \Pi(m_s,m_i)\Pi(n_s,n_i),
\end{equation}
where
\smallskip
\begin{align*}
\Pi(\mu,\nu)&=\frac{1}{\lambda^2z^2\sqrt{\pi
B_1}\mu!\nu!2^{\mu+\nu}}\sum_{k_1=0}^{\mu}\sum_{l_1=0}^{\nu}\sum_{k_3=0}^{\mu}\sum_{l_3=0}^{\nu}\mathcal{F}(\mu,\nu,k_1,l_1)\\&\times
\mathcal{F}^\ast(\mu,\nu,k_3,l_3)
\mathcal{K}(\mu+\nu-k_1-l_1,\mu+\nu-k_3-l_3),
\end{align*}
\begin{align*}\small
\mathcal{F}&(\mu,\nu,k,l) = \binom{\mu}{k}\binom{\nu}{l} 2^{\mu+\nu}i^{k+l}\sigma(k,l)\,
\Gamma\left(\frac{k+l+1}{2}\right)\\&\times
\left(\frac{\sqrt{2}}{W}\right)^{\mu+\nu-k-l}
\sqrt{1-\zeta}\left(\sqrt{\zeta}\right)^{k+l}F\left(-k,-l;\frac{1-k-l}{2};\frac{1}{2\zeta}\right),
\end{align*}
%\begin{strip}
%\begin{center}
%\resizebox{.4\hsize}{!}{
\begin{align*}
&\mathcal{K}(\mu,\nu)=\frac{1}{4}\left(\frac{1}{\sqrt{2}}\right)^{\mu+\nu} \sum_{p=0}^{\mu}
 \sum_{q=0}^{\nu}\binom{\mu}{p}\binom{\nu}{q}\\&\times
 (-1)^{\nu-q}\left(\frac{1}{\sqrt{C_1}}\right)^{2+p+q}
 \left(\frac{1}{\sqrt{C_2}}\right)^{\mu+\nu-p-q}\\&\times
\Biggl(%\left\{
  \sigma(0,p+q)\sigma(0,\mu+\nu-p-q) \sqrt{\frac{C_1}{C_2}}\
 \Gamma\left(\frac{1+p+q}{2}\right)%\right.
 \\&\times
\Gamma\left(\frac{1+\mu+\nu-p-q}{2}\right)F\left(\frac{1+p+q}{2};\frac{1+\mu+\nu-p-q}{2},\frac{1}{2};C_4\right)\\&-
\frac{i\sigma(1,p+q)\sigma(1,\mu+\nu-p-q)(4C_1C_2+C_3^2)}{C_2C_3(1+p+q)(1+\mu+\nu-p-q)}
\Gamma\left(\frac{2+p+q}{2}\right)
\\&\times
\Gamma\left(\frac{2+\mu+\nu-p-q}{2}\right)F\left(\frac{2+p+q}{2};\frac{2+\mu+\nu-p-q}{2},-\frac{1}{2};C_4\right)\\&+
\frac{i\sigma(1,p+q)\sigma(1,\mu+\nu-p-q)[4C_1C_2+C_3^2(4+\mu+\nu)]}{C_2C_3(1+p+q)(1+\mu+\nu-p-q)}\\&\times
\Gamma\left(\frac{2+p+q}{2}\right)%\left.
\Gamma\left(\frac{2+\mu+\nu-p-q}{2}\right)\\&\times
F\left(\frac{2+p+q}{2};\frac{2+\mu+\nu-p-q}{2},
\frac{1}{2};C_4\right)
\Biggr),%\right\},
\end{align*}
%\end{center}
%\end{strip}%
$F$ is the Hypergeometric function, $\binom{\cdot}{\cdot}$ is the binomial coefficient and $\Gamma$ is the Gamma function.
Other quantities are defined as
%\begin{equation}\nonumber
%\begin{split}
$\sigma(k,l)=(-1)^k+(-1)^{l}, \quad
\zeta=\frac{1+\Lambda_0^2}{1+\Lambda_0^2+i\Lambda_0},\quad
\Lambda_0 =\frac{2z}{kW_0^2}, \quad
W_0=\sqrt{2}W_{0p},\quad C_1=\mathrm{Re}A_2-\frac{A_3}{2},\quad
C_2=\mathrm{Re}A_2+\frac{A_3}{2},\quad
%\end{split}
%\end{equation}\nonumber
%\begin{equation}
%\begin{split}
C_3=\mathrm{Im}A_2,\quad C_4=-\frac{C_3^2}{4C_1C_2},\quad
A_1=\frac{k}{4z}\left(\frac{\Lambda_0}{1+\Lambda_0^2}-i\right),\quad
%\end{split}
%\end{equation}\nonumber
%\begin{equation}
%\begin{split}
A_{2}=-\frac{B_2^2}{4B_1}+B_3+\frac{k}{z}\frac{\Lambda_0}{1+\Lambda_0^2},\quad
A_3=-\frac{|B_2|^2}{2B_1}+B_4, \quad
B_1=\frac{k}{z}\left(\frac{1}{2\Lambda_0}+\frac{\Lambda_0}{2}+\gamma\right),\quad
B_2=\frac{k}{z}\left(\frac{1}{\Lambda_0}-\gamma -i\right),\quad
%\end{split}
%\end{equation}
%\begin{equation}
%\begin{split}
B_3=\frac{k}{z}\left(\frac{1}{2\Lambda_0}+\gamma-i\right),\quad
B_4=\frac{k}{z}\left(\frac{1}{\Lambda_0}+2\gamma\right).$
%\end{split}
%\end{equation}\nonumber
Here, $\gamma =1.63 (\sigma_R^2)^{\frac{6}{5}}$ measures the strength of the turbulence related to the Rytov variance,
\begin{equation}
\sigma_R^2=1.23 C_n^2k^{7/6}z^{11/6},
\end{equation}
where $C_n^2$ is the structure constant of the refractive index of the atmosphere, $k$ is the wavenumber.
\newline
The ensemble averaging in Eq. (\ref{prob_trans_modes1}) is performed using the Wiener-Khinchin Theorem with the Tatarskii power spectrum of index of refraction fluctuations. Similar calculation can be found in Ref. \cite{Avetisyan2016}.
Being one of the main results of our paper, expression (\ref{final_expression}) shows that the joint two-mode
detection probability for signal and idler photons
is a product of functions, which mixes the indices of signal and idler. This directly implies that the entanglement is preserved.
As the two-photon wavefunction in Eq.\,(\ref{2-phot-wf}) is expressed in
the paraxial approximation, thus not properly normalized, the sums $\sum_{m_s}\sum_{n_s}\sum_{m_i}\sum_{n_i}P(HG_{m_sn_s}, HG_{m_in_i})$
do not converge to unity. This is because the shapes of the higher and higher order modes increase, leading to a deviation from the
paraxial approximation.
\newline
To have better insight, one can arrange the values of Eq.(\ref{final_expression}) in a matrix. Below we construct the first 100 values of
(\ref{final_expression}) in a $10\times 10$ matrix for the vacuum case 
\\
\begin{widetext}
%\begin{center}
\smallskip
\begin{equation}
\left (
\begin {array} {ccccccccccc}
0.31307 & 0 & 0 & 0.03986 & 0 &
0.03986 & 0 & 0 & 0 & 0 \\
0 & 0.07697 & 0 & 0 & 0 & 0 &
0.02940 & 0 & 0.00980 & 0 \\
0 & 0 & 0.07697 & 0 & 0 & 0 &
0 & 0.00980 & 0 & 0.02940 \\
0.03986 & 0 & 0 & 0.04345 &
0  & 0.00508 & 0 & 0 & 0 & 0 \\
0 & 0 & 0 & 0 &
0.01892 & 0 &  0 & 0 & 0 & 0 \\
0.03986 & 0 & 0 & 0.00508 & 0 &
0.04345  & 0 & 0 & 0 & 0 \\
0 & 0.02940 & 0 & 0 & 0 & 0 &
0.03023  & 0 & 0.00374 & 0 \\
0 & 0 & 0.00980 & 0 & 0 & 0 & 0 &
0.01068 & 0 & 0.00374  \\
0 & 0.00980 & 0 & 0 & 0 & 0 &
0.00374 & 0 & 0.01068 & 0 \\
0 & 0 & 0.02940 & 0 & 0 & 0 &
0 & 0.00374 & 0 & 0.03023 \\
\end {array}
\right)
\end{equation}
\end{widetext}
\\
\newline
The  elements $ij=m_sn_s,m_in_i$ of  the  matrix  are
double indices corresponding to mode numbers of signal
and idler ranging as $m_kn_k= \lbrace
00$,$01$,$10$,$02$,$11$,$20$,$03$,$12$,$21$,$30
\rbrace $, where $k=s,i$.
Moreover, the matrix elements satisfy selection rules obtained in Ref. \cite{Walborn2005}
\begin{align}\label{selection-rules}
\mathrm{parity}(m_s+m_i) &= \mathrm{parity}(m_p), \quad m_s+m_i \geq m_p,\\
\mathrm{parity}(n_s+n_i) &= \mathrm{parity}(n_p), \quad n_s+n_i \geq n_p.
\end{align}
The matrix for a weak turbulence regime, more precisely, for $\sigma_R^2=0.02$ and for the propagation distance $z=5$km, the wavelength $\lambda=0.8\mu$m and the spot size of the pump at the nonlinear crystal $W_0=10$ cm has the form shown in Eq.(\ref{matr-turb}), . We see that all elements are different from zero: the atmosphere causes crosstalk between different modes. The variation of the first two matrix elements with the strength of turbulence is shown in Fig.\ref{fig1}.
\newline
For the first line of the matrix, we compare the behavior of the probabilities for two different turbulence conditions $\sigma^2_R=0.01$, and $\sigma^2_R=0.1$ with the vacuum case.
We let the propagation distance, the Fresnel ratio and the spot size of the pump at the crystal to be the same. One can see that the crosstalk
between modes is not uniform: photons tend to stay in some modes, e.g., \{00,02\} and \{00,20\}, conversely,
crosstalk to some modes, e.g., \{00,01\}, \{00,10\} is more preferred than to others, e.g., \{00,12\} and \{00,21\}.
Therefore, in making quantum communication with HG alphabet, one has a distinctive choice of modes that can increase the fidelity of the communication. One should also note that this is true for quite weak turbulence conditions as demonstrated in Fig.\ref{Fig2}.
\smallskip
\begin{figure}
\centering
\subcaptionbox{\label{a1}}{
  \includegraphics[height=0.2\textheight, width=0.4\textwidth]{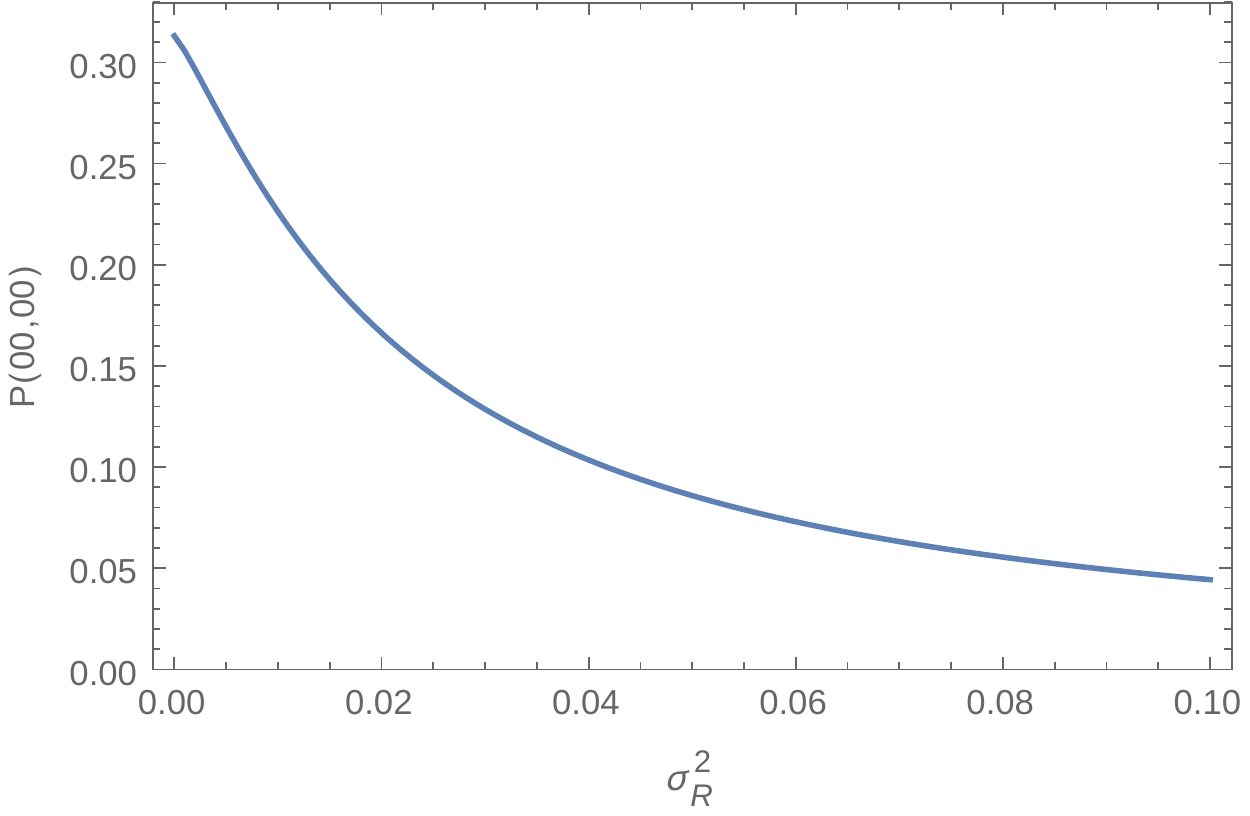}
  }\par\medskip
\subcaptionbox{\label{b1}}{
\includegraphics[height=0.2\textheight,width=0.4\textwidth]{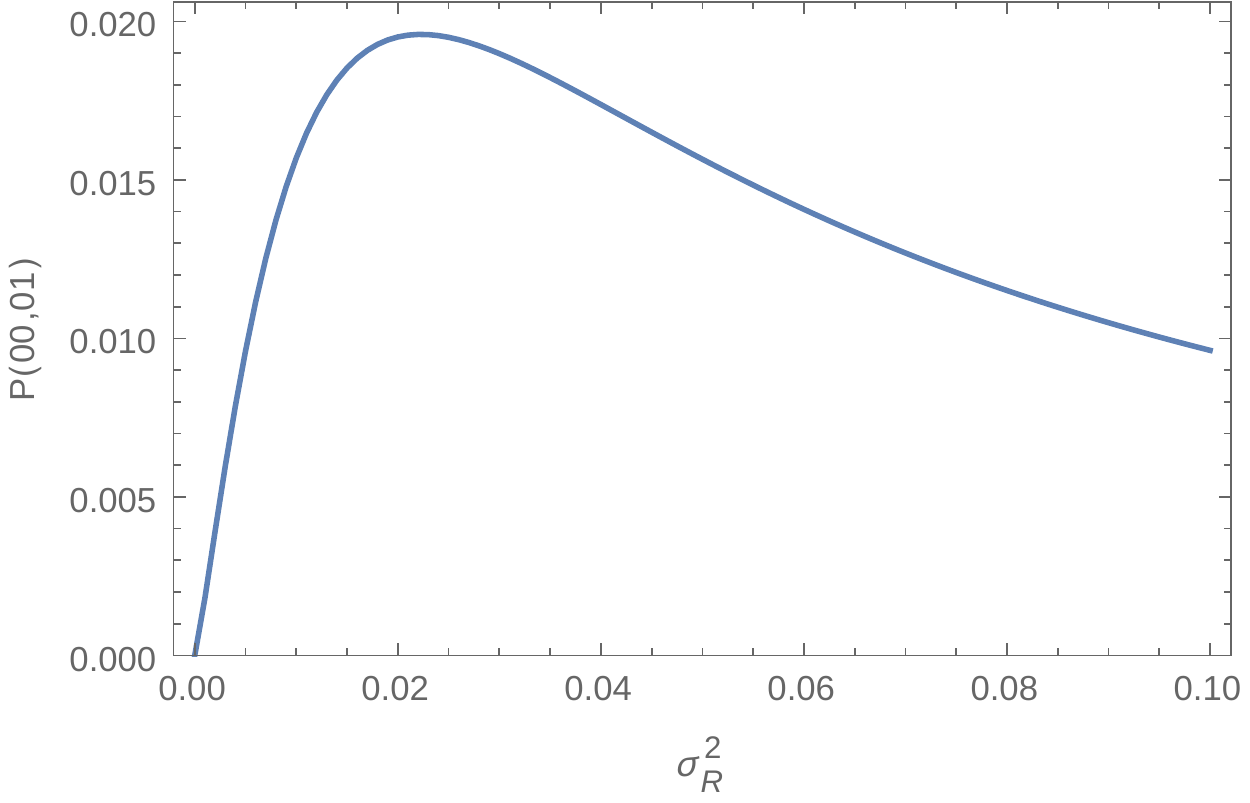}
}\par\medskip
  \caption{The behaviour of the two-mode joint probabilities $P(00,00)$ (\ref{a1}) and $P(00,01)$ (\ref{b1}) as turbulence strength increased.}
  \label{fig1}
  \end{figure}
\begin{widetext}
\begin{center}
\smallskip
\begin{equation}\label{matr-turb}
\left (
\begin {array} {cccccccccc} 0.2262 & 0.0157 & 0.0157 & \
0.0379 & 0.0011 & 0.0379 & 0.0077 & 0.0026 & \
0.0026 & 0.0077 \\ 0.0157 & 0.0439 & 0.0011 & \
0.0009 & 0.0030 & 0.0026 & 0.0204 & 0.0001 & \
0.0073 & 0.0005 \\ 0.0157 & 0.0011 & 0.0439 & \
0.0026 & 0.0030 & 0.0009 & 0.0005 & 0.0073 & \
0.0001 & 0.0204 \\ 0.0379 & 0.0009 & 0.0026 & \
0.0275 & 0.0001 & 0.0063 & 0.0005 & 0.0019 & \
0.0001 & 0.0013 \\ 0.0011 & 0.0030 & 0.0030 & \
0.0001 & 0.0085 & 0.0001 & 0.0014 & 0.0002 & \
0.0002 & 0.0014 \\ 0.0379 & 0.0026 & 0.0009 & \
0.0063 & 0.0001 & 0.0275 & 0.0013 & 0.0001 & \
0.0019 & 0.0005 \\ 0.0077 & 0.0204 & 0.0005 & \
0.0005 & 0.0014 & 0.0013 & 0.0191 & 0.0001 & \
0.0034 & 0.0003 \\ 0.0026 & 0.0001 & 0.0073 & \
0.0019 & 0.0002 & 0.0001 & 0.0001 & 0.0053 &\text\
3\times10^{-6} & 0.0034 \\ 0.0026 & 0.0073 & 0.0001 & \
0.0001 & 0.0002 & 0.0019 & 0.0034 &\text \
3\times 10^{-6} & 0.0053 & 0.0001 \\ 0.0077 & 0.0005 & \
0.0204 & 0.0013 & 0.0014 & 0.0005 & 0.0003 & \
0.0034 & 0.0001 & 0.0191 \\
\end {array} \right)
\end{equation}
\end{center}
\end{widetext}
\begin{figure}[H]
 \centering
 \subcaptionbox{\label{a2}}{
\includegraphics[height=.155\textheight, width=.35\textwidth]{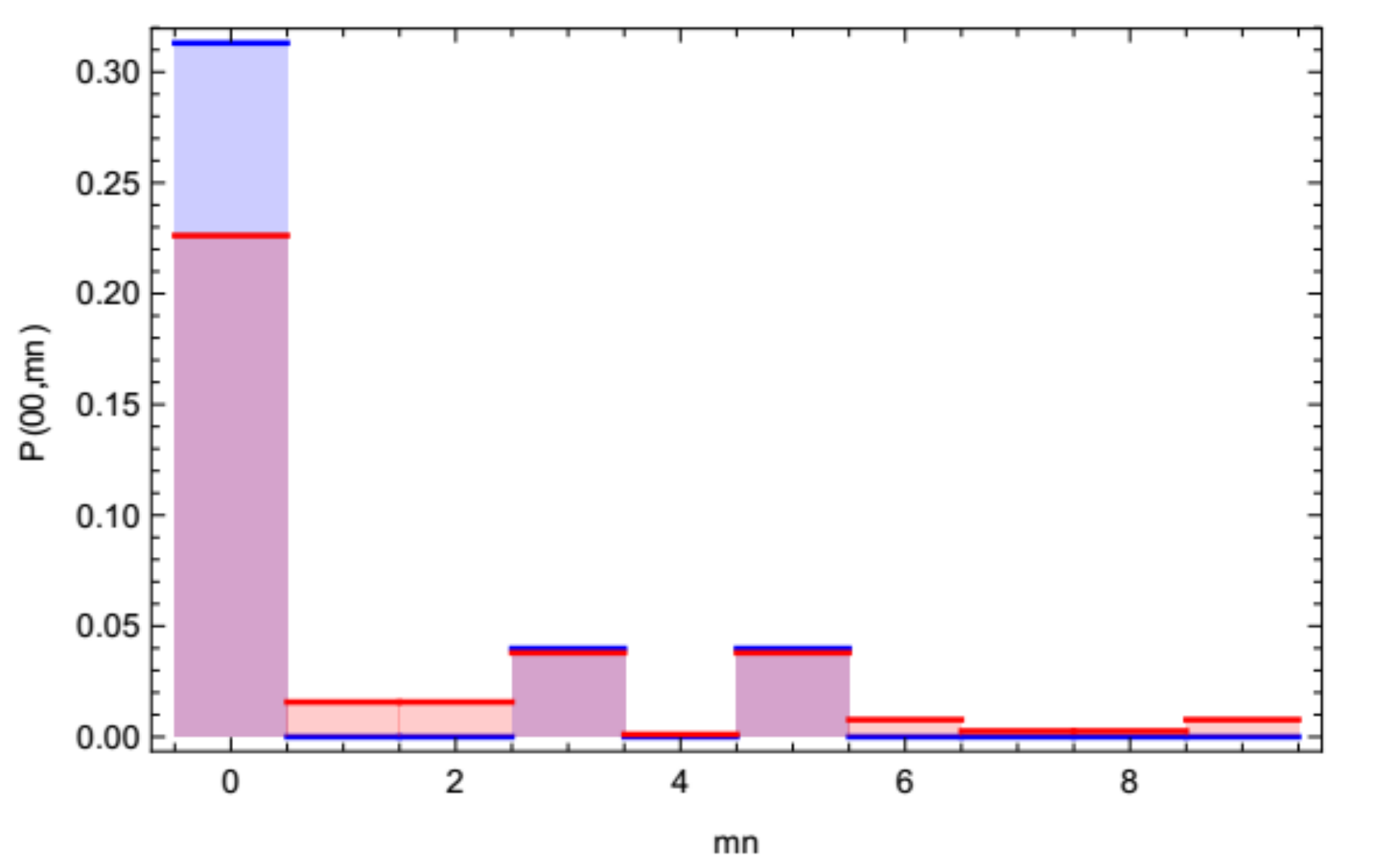}
 }\par\medskip
\subcaptionbox{\label{b2}}{
\includegraphics[height=.155\textheight,width=.35\textwidth]{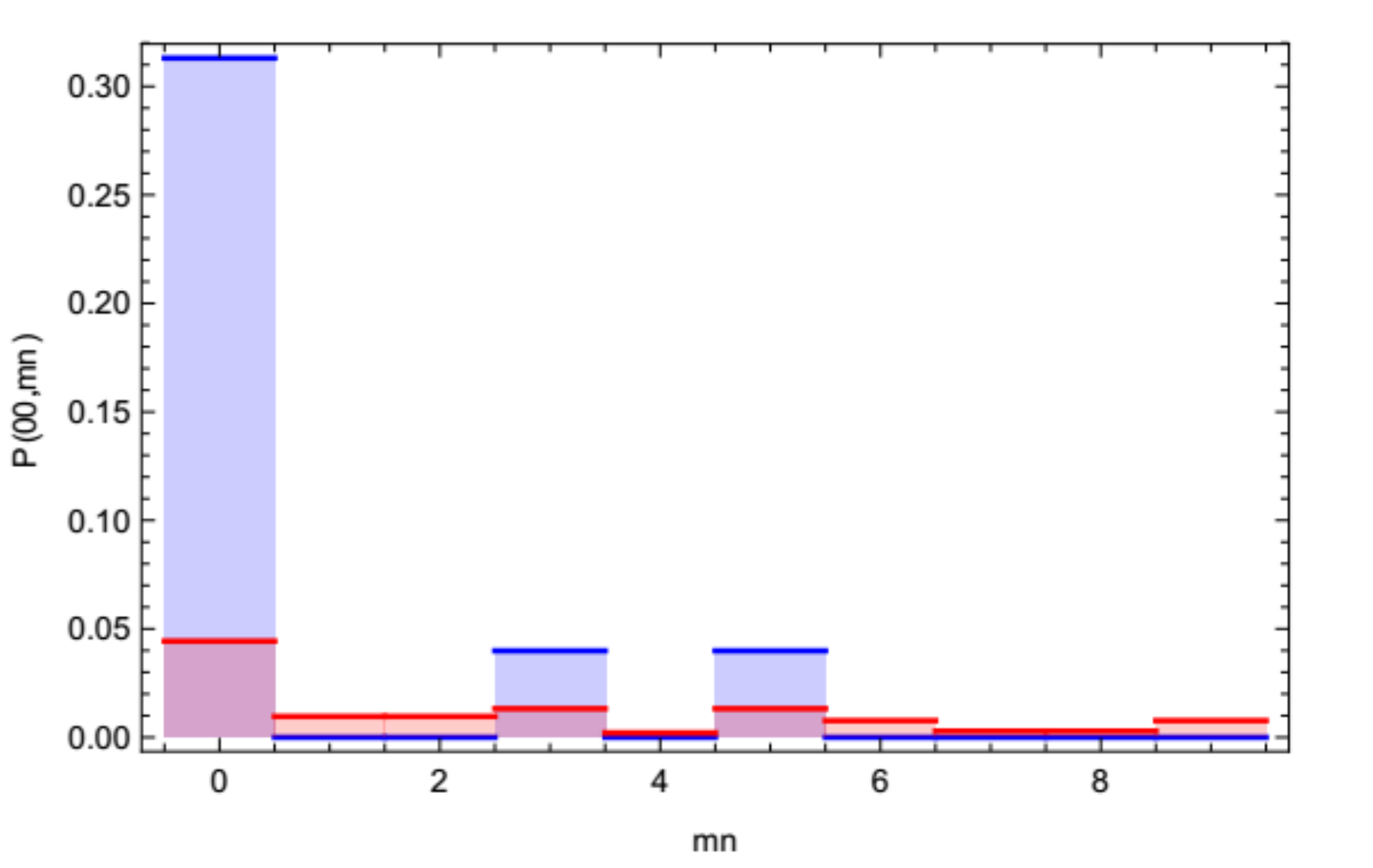}
}\par\medskip
\caption{Comparison of the probabilities in the first lines of the two matrices for turbulence strengths $\sigma_R^2=0,01$ (\ref{a2}) and $\sigma_R^2=0,1$ (\ref{b2}).}
\label{Fig2}
\end{figure}
Fig.\ref{fig1} shows that forbidden probabilities, imposed by the selection rules (\ref{selection-rules}), increase due to the crosstalk between modes caused by atmosphere.  Accordingly, the allowed probabilities decrease to conserve the total probability.
In Fig.\ref{Fig2}, the blue and red columns represent the probabilities for the vacuum and turbulent cases, respectively.
\section{Conclusion}
The quantum state produced by SPDC process is entangled in spatial degrees of freedom. The entanglement
in the HG transverse modes has been shown by Walborn \textit{et} \textit{al.} \cite{Walborn2005} by implying
restrictions on both the parity and order of the down-converted HG fields, recapped in Eq.(\ref{selection-rules}).
One could use this higher dimensional entanglement to make quantum communication with a large alphabet, thereby enhancing
the security of the communication. For a long distance and, eventually, global quantum communication with entangled photonic states one needs
to consider the effects of the atmosphere. We obtained an analytic expression for the joint detection probability for
signal and idler photons either of them to be found in an HG mode of any order. We considered a Gaussian beam as a pump and used the paraxial
approximation for the down-converted fields. Our results show that for a propagation distance of 5km there is a nonuniform crosstalk
between modes: there are modes that tend to stay populated while some tend to stay empty. This feature can be used to enhance the
quantum communication fidelity by selecting appropriate mode projectors at the detectors' side.

%\subsection{Sample Dataset Citation}
%1. M. Partridge, "Spectra evolution during coating," figshare (2014) [retrieved 13 May 2015], http://dx.doi.org/10.6084/m9.figshare.1004612.

%\section{Funding Information}
%National Science Foundation (NSF) (1263236, 0968895, 1102301); The 863 Program (2013AA014402).
%The authors thank H. Haase, C. Wiede, and J. Gabler for technical support.
%Do not use Funding Information or Acknowledgment headings.

%\section{References}
%Note that \emph{Optics Letters} uses an abbreviated reference style. Citations to journal articles should omit the article title and final page number; this abbreviated reference style is produced automatically when the \emph{Optics Letters} journal option is selected in the template, if you are using a .bib file for your references.

%However, full references (to aid the editor and reviewers) must be included as well on a fifth informational page that will not count against page length; again this will be produced automatically if you are using a .bib file.

%\bigskip
%\noindent Add citations manually or use BibTeX. See \cite{Zhang:14,OSA,FORSTER2007}.

% Bibliography
%\bibliography{sample}

% Full bibliography added automatically for Optics Letters submissions
% Note that this extra page will not count against page length
%\ifthenelse{\equal{\journalref}{ol}}{
%
%\clearpage
%\bibliographyfullrefs{sample}
%}{}

% Please include bios and photos of all authors for aop articles
%\ifthenelse{\equal{\journalref}{aop}}{%

\end{document}